# A New Lens on Multi-GPU Communication: API-Level Memory Models for Portable Optimization

Johnathan Alsop, Bradford Beckmann, Brandon Potter, Jiakun Yan, Aurelien Bouteiller,
Omri Mor, Tony Tye, Sameer Sahasrabuddhe
AMD

*Abstract*— As network communication increasingly becomes a performance bottleneck for ML and HPC workloads, it is more important than ever to optimize multi-GPU data transfers. Recently, low-level device APIs have emerged that allow programmers to perform fine-grain optimizations via one-sided put/get operations triggered directly from the device. However, the communication libraries providing these device APIs (e.g., rocSHMEM, MSCCL++, and NCCL/RCCL GIN) currently lack a rigorous memory model defining how their primitives interact with the underlying memory systems. Without this, emerging HPC and ML workloads aggressively optimizing network communication via device APIs can be brittle, error-prone, and face compatibility and inter-operability challenges.

This paper proposes the use of API-level memory models for these emerging device communication APIs. An API-level memory model defines the memory semantics of communication API calls in a way that *abstracts* lower-level details from the API user and simplifies development. This work defines one such model for device-side communication (implemented in AMD's recently open sourced rocSHMEM library) and demonstrates how formal relaxations to such a model can enable low-level LLM performance optimizations while retaining a well-defined standardized interface. Specifically, we describe how updating the SHMEM API implemented by rocSHMEM can properly support cache bypassing for network data, improving performance of a collective communication workload by up to 2.48x, a fused compute+communication workload by up to 1.77x, and a state-of-the-art expert-parallel communication workload by up to 1.18x.

## I. Introduction

The rise of deep learning and large language models (LLMs) has led to powerful AI models with billions to trillions of parameters. The recent growth of these models and the associated explosion of memory and compute demands motivate the use of large networks of compute devices for performing training and inference. Scaling across vast numbers of devices requires utilizing many forms of ML parallelism and frequent communication and synchronization between devices. Furthermore, as HPC and ML applications scale to an ever increasing number of compute devices, communication is becoming an even greater performance bottleneck.

Distributed applications commonly leverage communication libraries with standardized APIs to simplify programming and hide the low-level details of inter-device data transfer. These APIs generally offer an easily addressable partitioned global address space and one-sided asynchronous communication primitives like `put` and `get` that avoid the performance overheads of blocking for long-latency network operations. Finally, specialized ordering operations enable flexible and targeted synchronization of these asynchronous primitives. Standardized APIs like OpenSHMEM [9] (abbreviated here as SHMEM) and MPI [10] have supported these properties for decades in the CPU domain, and they are increasingly used (or will soon be adopted) in industry leading GPU ML libraries [1][11] [12][28][29][30][31].

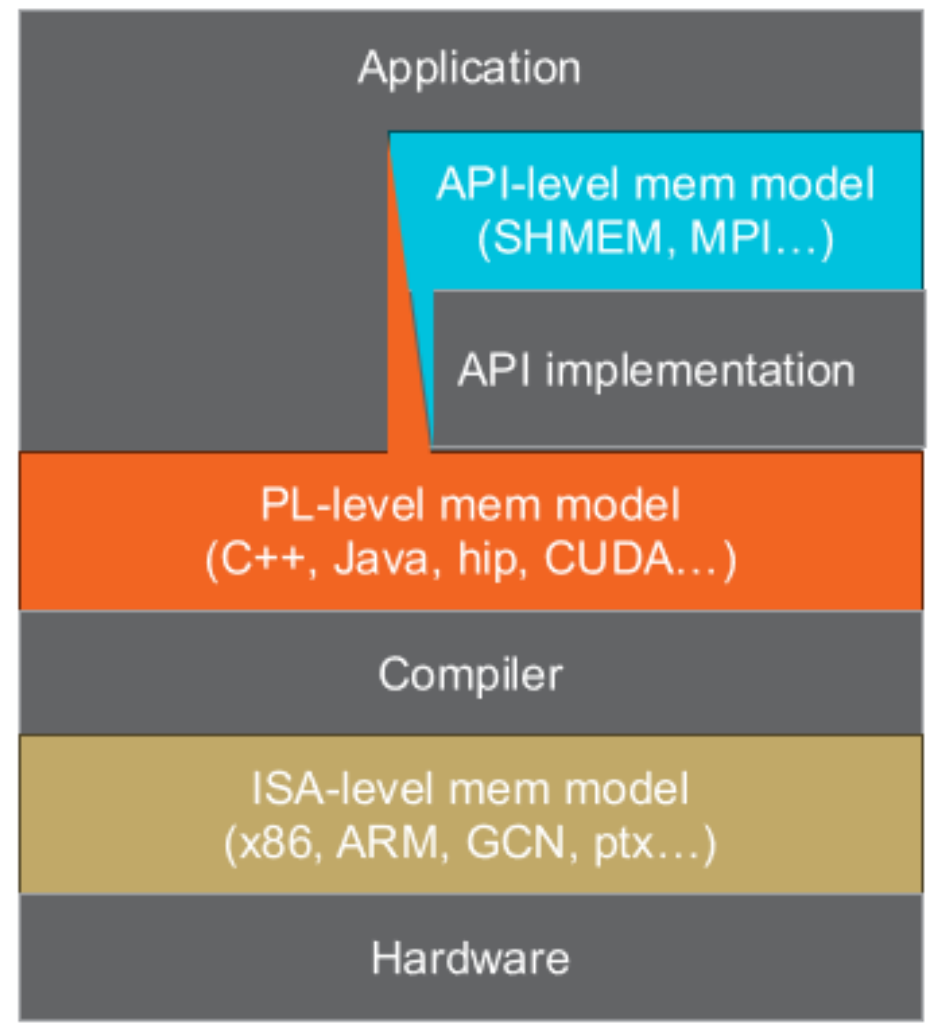

**Figure 1: The API-level memory model (in blue) builds upon the PL-level memory model and defines the interface between application software and the API implementation.**

Standardized APIs support cross-company innovation. They enable inter-operability of applications with a multiplicity of vendor implementations, and they limit complexity through a separation of concerns. This allows both implementers and users of an API to safely and independently introduce optimizations into their code, as long as they respect the API guarantees. However, the way current communication APIs define these guarantees has been imprecise, particularly for non-CPU systems. Communication operations can require complex network-dependent interactions between software threads and/or dedicated hardware units. In addition, GPU threads communicating through memory often require different and more complex coherency characteristics than CPU-only systems. With both GPU hardware and software evolving rapidly, imprecise specification complicates the creation of performant and portable code. We therefore need formal specifications of memory ordering to enable aggressive optimization on both sides of the API while ensuring correctness.

To this end, we propose the use of an API-level memory model for communication, and we define one such model specifically for SHMEM. An API-level model is a contract between an application developer and an API implementer that specifies how the API interacts with the underlying memory system. Figure 1 reflects the API-level memory model's position in the software stack. In contrast with existing memory models that describe interactions at the level of the Instruction Set Architecture (ISA) or Programming Language (PL), an API-level model operates on higher-level API operations (put, get, barrier, etc.). In GPU device-side communication libraries, these operations can be coarse-grain, paired with other operations, and/or asynchronous. We formally integrate these properties into an underlying PL-level memory model by 1) defining the observable accesses of each API operation, 2) fusing paired operations into a single event, and 3) breaking program order for these observable accesses and defining new relations to enforce API-specific ordering constraints.

In addition to improving programmability, a formal API-level memory model provides a basis for optimized variants of the API. We describe one such variant and we detail the semantic extensions needed for GPU device-side communication APIs to enable this optimization. Once again, formal specification is necessary here so that API users and developers know and agree upon which ordering constraints will be preserved, weakened, or strengthened.

This work makes the following contributions:

1. We identify limitations in existing network communication API specifications and introduce the notion of a formal API-level memory model to address them.
2. We define and validate an axiomatic API-level memory model specifically for the SHMEM API that leverages and extends familiar consistency relations and semantics to specify memory ordering for API operations and their accesses.
3. We describe how the existing SHMEM and RCCL/NCCL GIN APIs and our newly defined API-level models can be extended to enable a communication optimization: targeted ordering for network data. We then demonstrate the performance benefits of this optimization for some important communication patterns, showing speedups of up to 2.48x.

## II. Background

### A. Memory Consistency Models

A Memory Consistency Model (MCM) defines what values loads are allowed to return in a multi-threaded program. Today's MCMs are frequently defined at the ISA-level (e.g., TSO [2] and PTX [3]), or the PL-level (e.g., C++ [14], Java [15], or HSA [16]). These represent a contract between layers of the compute stack; they define allowable behavior for developers above the MCM layer, and theyrestrict legal reorderings by software and hardware below the MCM layers.

One common way to formally specify a MCM is via axioms over relations. Axiomatic formalization begins by enumerating all possible "candidate" executions of a parallel program. In defining these executions, loads may return the value of a conflicting store anywhere else in the execution, or the initial memory value at that address (i.e., allowing all possible reorderings). We then define relations between accesses in candidate executions, and based on properties over these relations, we define which executions are legal (restricting the reorderings that are permitted). The following sections rely on language and terminology used in past works to describe this formalism [13].

```
Sequential Consistency
sc = acyclic(po ∪ rf ∪ co ∪ fr)
Happens-before
hb = (po ∪ so)⁺
Data race
data_race = (a, b): a.addr == b.addr
  ∧ (a.type == ST ∨ b.type == ST)
  ∧ ¬ (a.atomic ∧ b.atomic)
  ∧ (a, b) ∉ hb ∧ (b, a) ∉ hb
```

**Figure 2: SC for DRF memory model formalism.**

*a) Sequential Consistency for Data Race Free (SC for DRF):* Modern systems use relaxed memory models that permit important reorderings at the microarchitecture and compiler levels. Although these are more complex than sequential consistency (i.e., consistent with an in-order interleaving of all thread accesses), or SC, this complexity is frequently mitigated by the use of the SC for DRF paradigm. As defined by Adve et al. [4], an SC for DRF model defines explicit synchronization semantics (e.g., acquire and release atomics) that must be used when communicating between threads to protect non-synchronizing "data" accesses. If proper synchronization is used, then a program is guaranteed to be SC.

Formalizing SC for DRF begins with defining SC axiomatically (as shown in Figure 2) based on a set of baseline relations: **program order (`po`)** relates all accesses within a single program thread based on the order they are evaluated; **reads-from (`rf`)** relates loads to the store that produced the returned value; **coherence order (`co`)** relates stores to the same address based on the order they arrive in memory; and **from-reads (`fr`)** relates reads with subsequent writes to the same address in `co` (defined as $rf^{-1};co$). **synchronization order (`so`)** relates accesses that establish a `rf` relationship across threads and legally synchronize by using acquire and release atomic semantics. **happens-before (`hb`)** leverages `so` and `po` to relate events that are guaranteed to happen in a specified order according to the memory model. **data race (`data_race`)** is defined as two conflicting data accesses where at least one access is a store, and they are not related by `hb` in a candidate execution. If all SC executions are free of data races, then the program will be SC. If data races exist, then behavior is undefined. Given the advantages of SC for DRF and its ubiquity in modern CPU and GPU systems, we assume the use of SC for DRF (or DRF for short) PL-level memory models going forward.

### B. GPU Coherence

GPUs use software-managed cache coherence to support high-throughput cache demands. This involves triggering bulk cache flush and invalidate actions for synchronization operations with release and acquire semantics, respectively, which can be expensive when GPU threads frequently synchronize. To enable more efficient GPU synchronization, modern GPU memory models use scoped synchronization, which limits how deep a flush or invalidate must be applied. This scoped variant of DRF is referred to as heterogeneous race free (HRF)[5], and it places further constraints on what is considered legal synchronization (acquire+release *and* matching synchronization scope).

**Table 1: A subset of SHMEM operations that impact the memory model, categorized as remote memory access routines (RMA), synchronization operations (SO), collective operations (Col), and ordering operations (Ord). Blocking indicates whether the operation is a blocking or non-blocking ("B" if there is only a blocking version, "B or NB" if there is a blocking and non-blocking version).**

| Name | Type | Blocking? | Description |
|---|---|---|---|
| put | RMA | B or NB | Transfer specified range of data from local memory to remote memory |
| get | RMA | B or NB | Transfer specified range of data from remote memory to local memory |
| atomic_fetch | SO | B or NB | Synchronizing read to remote memory |
| atomic_set | SO | B | Synchronizing write to remote memory |
| atomic_rmw | SO | B or NB | Synchronizing atomic read-modify-write operation on remote memory. Operations include swap, inc, add, and, or, xor. Different variants may or may not return a value. |
| put_signal | SO | B or NB | Put operation serialized with a subsequent synchronizing write to a signal variable in the same remote PE memory. |
| signal_set | SO | B | Synchronizing write operation to signal variable in remote memory |
| wait_until | SO | B | Block until a signal variable in local memory meets a specified condition |
| barrier | Col | B | Block until all other PEs in a team have reached a matching barrier operation and completed all outstanding nonblocking operations |
| alltoall | Col | B | Exchange a range of data blocks between all pairs of PEs in a team |
| broadcast | Col | B | Broadcast a block of data from a specified root PE to one or more destination PEs in a team |
| reduce | Col | B | Perform a reduction over a block of data in each PE in a team and store the result to all PEs. Different reduce operations are available, including sum, product, max, min, and, or, and xor. |
| fence | Ord | B | Order delivery of memory operations before and after this operation that target the same PE. |
| quiet | Ord | B | Block until all prior accesses and operations have completed. |

### C. *The SHMEM API*

While the ideas in this work can apply to most GPU device-side communication APIs, we focus on SHMEM [6][7] because of its widespread use in HPC and increasingly in ML frameworks. In this section we provide background on the SHMEM API with a focus on aspects relevant to memory consistency.

#### 1) *SHMEM Overview*

SHMEM defines a set of functions, or operations, that can set up the SHMEM environment, access memory, transfer data between processing elements (PEs) in a network, and/or order other operations. SHMEM relies on a symmetric heap for simplified addressing in a partitioned global address space (PGAS). Accesses only conflict if they target the same address and the same PE. Operations can additionally specify a context (ordering domain) and a team (a subset of PEs that can communicate).

#### 2) *SHMEM Operations*

A representative subset of MCM-relevant SHMEM operations are listed in Table 1 (see the OpenSHMEM specification [9] for the full set of operations and their variants). They are classified into four broad types**: remote memory access routines (RMA)** perform bulk data transfers between PEs; **synchronizing operations (SO)** can be used to access synchronization variables; **collective routines (Col)** are operations that span multiple PEs in a team, and **ordering operations (Ord)** order the visibility of SHMEM operations (discussed next). Many of these operations have blocking and non-blocking variants.

#### 3) *SHMEM Ordering Constraints*

Here we summarize and categorize existing SHMEM ordering semantics. Section IV identifies ambiguities in this specification, then formalizes the concepts, resolving the ambiguities.

**Implicit local visibility:** Although not explicitly specified by the API, local memory accesses from a PE prior to a SHMEM operation on the same PE should be visible to that operation. That means if a thread stores to a buffer, then issues a SHMEM put operation targeting that buffer, the prior buffer stores will be visible to the put operation.

**Local completion ordering:** For blocking remote memory access operations, the local accesses associated with the transfer (e.g., local reads for a put, local writes for a get) must "complete" before the API call "returns."

**Remote delivery ordering**: SHMEM specifies that a `fence` orders the "delivery" of SHMEM operations from the same context at a remote PE, if those operations are separated by `fence` and target the same PE. `fence` also orders prior memory accesses with subsequent remote SHMEM operations. However, it does not order prior operations with subsequent memory accesses and it does not order operations to different PEs.

**Remote completion ordering**: `barrier` and `quiet` operations go further than `fence` in that they do not "return" until all prior SHMEM operations in the same context and their associated (local and remote) accesses have "completed."

**Synchronization requirements**: Only the synchronizing and collective operations defined above can be used to synchronize between PEs (not RMAs). In addition, synchronizing operations must use a compatible context and team, and every dynamic collective operation must be matched with a corresponding call issued in the same order in each member of the associated team. If these rules are violated, behavior is undefined.

## III. Related Work

API memory models build upon a large body of past memory consistency formalization research [2][3][4][5][13][14][15][16][19][20][22][23][24]. These past efforts formalize the semantics of existing weak memory models, and, in some cases, they prove the correctness of compilation strategies or verify the legal behavior for relevant code patterns (litmus tests). However, their focus on PL-level and ISA-level semantics (semantics of loads, stores, and low-level synchronization actions) sets them apart from this work, which focuses on the semantics of higher level SHMEM operations (e.g., put, get, reduction, etc.) This difference presents new formalization challenges when integrating API semantics with a PL-level model, and it complicates similar validation efforts for compilation (there are many correct ways to "lower" an API operation to PL-level) and litmus tests (correct behavior for such tests has not previously been formally defined).

Batty et al. define a methodology for abstracting library semantics for C programs into high-level *guarantee* and *deny* relations over a history of call and return events [21]. This provides behavioral guarantees for library users without requiring them to understand library internals, similar to an API-level model. However, this methodology is designed to be general enough for any potential library. As such, it places limits on library usage (a non-interference requirement restricts user and library code to communication solely through call and return semantics) and it prevents the simpler and more targeted definitions of ordering semantics and optimizations that are possible when considering a specific library (e.g., SHMEM).

Recent work by Ambal et al. [25] and Dan et al. [26] formalizes the behavior of remote direct memory access (RDMA) put and get operations on CPU devices. Similar to SHMEM put and get operations, these require integrating a notion of asynchrony into the memory model. However, the semantics are tied to stronger (SC and TSO) CPU memory models and low-level RDMA implementation details (queue pairs, request and response buffers, poll operations, etc.), limiting their portability to more relaxed systems (e.g., GPUs) and different networking technologies. In contrast, the API-level model described here integrates with a relaxed SC for DRF model and presents a higher level SHMEM interface that offers more flexibility to both users and implementers.

Recently, NVIDIA introduced the notion of memory synchronization domains to reduce the cost of synchronization actions [17]. A synchronization domain is an ID associated with each GPU kernel launch. When enabled, only cached data that was accessed from the matching synchronization domain needs to be flushed/invalidated on a synchronization event. This can limit the impact of synchronization when there are multiple concurrent kernels running on a GPU, and it has a minimal impact on the SC for HRF memory model. However, synchronization domains do not reduce the overheads when only a single kernel is active, and they require the use of system-scope for any inter-kernel synchronization.

Finally Hamidouche et al. [31] introduced GPU Initiated Networking (GIN) APIs into NCCL that were quickly adopted by AMD's RCCL [35]. The current synchronization rules defined in GIN exhibit ambiguity similar to that of SHMEM, and they disallow the targeting ordering optimizations described in this paper.

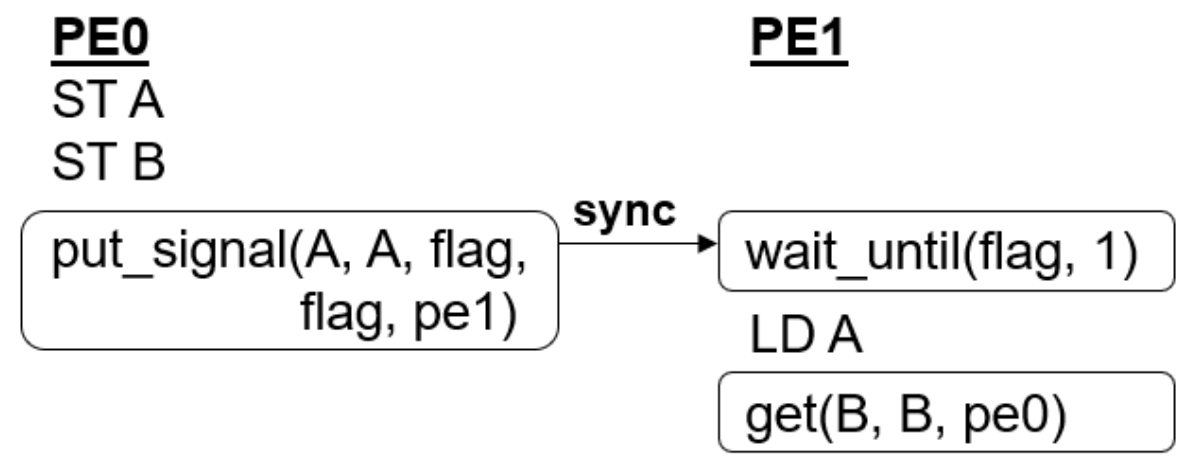

**Figure 3: Example SHMEM code with PEs synchronizing via put_signal and wait_until operations.**

## IV. The API-Level Memory Model

An API-level memory model represents a contract between the implementer of an API and an application programmer using the interface. We first describe how the existing SHMEM specification leaves certain ordering questions ambiguous. We then provide these missing details by defining an axiomatic API-level SHMEM memory model. Finally, we describe some intuitive rules for enforcing the constraints of the formal model.

### A. Limits of the SHMEM Specification

The SHMEM specification summarized above uses CPU-centric language to define the ordering constraints of SHMEM programs. It describes when an API call may "return" relative to the "delivery" or "completion" of various memory accesses and transfers, but these terms are not precisely defined. This ambiguity can lead to performance and correctness challenges on emerging systems. To illustrate these challenges, consider the program depicted in Figure 3. PE0 performs multiple accesses to local memory, then issues a `put_signal` operation that transfers some of the accessed data to PE1. PE1 synchronizes with the signal using a paired `wait_until` operation, then accesses the transferred data and issues a `get` targeting the remaining data previously stored on PE0.

When implementing this code on a GPU or other accelerator with software-managed coherence, data visibility between application code, API code, and communicating PEs must be explicitly managed. PL-level acquire and release semantics (triggering cache flush and invalidate operations) may be needed to make the store of `A` visible to the load in `put_signal` (i.e., if the `put_signal` is implemented in a separate thread), to avoid reading stale cached data after the `wait_until` operation, and to ensure the store of `B` is visible to the agent implementing the `get` operation from PE1. Language in the current SHMEM specification does not clarify who is responsible for these actions, or even whether the store of `B` must be visible to the `get`; should the application developer insert a release fence before the `put_signal` to ensure prior stores are visible? Do they need to insert an acquire fence after the `wait_until` if stale data may exist in the cache? Should these actions be handled by the API? Which data exactly will be made visible? When API users and developers answer these questions differently, it can result in errors (due to not enough cache actions) or inefficiency (due to excessive cache actions).

**Table 2: Minimal observable accesses for a subset of example SHMEM operations.**

| Operation | Observable Accesses |
|---|---|
| get(dst, src, size, src_pe) | `array = LD src[0:size-1], src_pe`<br>`ST array -> dst[0:size-1], my_pe` |
| put(dst, src, size, dst_pe) | `array = LD src[0:size-1], my_pe`<br>`ST array -> dst[0:size-1], dst_pe` |
| wait_until(signal, value) | `while (LD signal, my_pe != value) ;` |
| fence, quiet, barrier, sync | `<No observable memory accesses>` |

Beyond the question of who is responsible for cache actions in SW-managed devices, application programmers need clarity on how the API interacts with the PL-level memory model. For example, we may extend the example code in Figure 3 by adding a SHMEM `atomic_set` operation targeting PE0 at the end of the existing PE1 code. If the set value is read by a normal atomic access in a *second* thread on PE0, is this a legal synchronization? Would it establish a `hb`-ordering between the two threads on PE0? The answers to these questions are not explicitly addressed by the current SHMEM specification.

The issues and code patterns described here may not be common in current distributed CPU SHMEM applications. However, as domains like distributed ML increasingly motivate the use of accelerators like GPUs and aggressive network communication optimizations, formalization becomes more important. In the following sections, we formalize how an API like SHMEM interacts with a PL-level memory model. In doing so, we aim to resolve this uncertainty and provide a basis for further refinements of the API memory model.

## *B. Defining an API-Level Memory Model*

Here we define an API-level memory model that builds on top of an underlying PL-level memory model while formalizing the ordering constraints required by the API specification (or defining them where ambiguous). For simplicity, we focus on the basic DRF PL-level model described in Section II.A, but the integration proceeds identically for consistency models with relaxed atomics or scoped synchronization.

A SHMEM program contains API operations alongside memory accesses within each thread. SHMEM operations are considered events ordered by `po` just like any other access. However, the semantics of these operations are more complex than the loads, stores and sync operations that PL-level models operate on. SHMEM operations have three general properties that must be handled for PL-model integration: 1) coarse granularity, 2) multi-PE collective operations, and 3) explicit asynchrony. We address these challenges by 1) defining a minimal set of observable accesses for each operation, 2) fusing collective multi-PE operations, and 3) breaking program order with the calling thread and introducing new relations into the memory model that enforce the API's ordering semantics. These new events and relations are then used to extend the definitions of `hb`, `data_race`, and `sc` in an API-level DRF memory model.

### *1) Minimal Observable Memory Accesses*

We first define how operations access memory. For the purposes of the API-level SHMEM memory model, we only consider accesses that are observable to the SHMEM user. This is a minimal set that does not include accesses to intermediate buffers and synchronization structures, which are implementation specific and should not be user-visible. Table 2 provides the minimal observable memory accesses for a subset of SHMEM operations. SHMEM operations can access data in local and remote PEs, so observable accesses must also specify the PE ID. Note that some operations access multiple memory locations (coarse-grain operations), and some do not have any observable accesses (ordering-only operations) and are mainly used for synchronization and constraining asynchrony. For coarse-grain remote transfer operations where the source and destination targets are simply specified as a range of addresses (put, get), the full range of addresses are assumed to be accessed in a single event.

### *2) Fusing Multi-PE Operations*

SHMEM and similar communication APIs have collective operations that synchronize across multiple PEs in a team, in some cases without any observable synchronizing accesses (e.g., barrier). We merge these operations into a single event for the purposes of the memory model. This single operation and any associated observable accesses are considered to be executed collectively by all nodes in the synchronizing set of PEs, allowing us to establish ordering relations across these PEs.

### *3) Relations on SHMEM Operations and Accesses*

To support asynchronous (non-blocking) operations in SHMEM, *the observable accesses associated with a SHMEM operation are NOT related by* `po` *with accesses from the calling thread*. Conceptually, it is as if each operation issues its accesses from a separate thread (which is consistent with how they will often be implemented). A `po` relationship is, however, applied between observable accesses *within* the SHMEM operation, but only if 1) they exhibit a true data or control dependency, or 2) the API specifically constrains reordering (this is the case for `put_signal`, where all `put` accesses are `po` ordered with the final `signal` access). Since the `po` relation does *NOT* exist between an operation's observable accesses and normal memory accesses in the calling thread, these accesses can form a data race with each other.

We define a set of relations in Figure 4 that constrain the ordering of SHMEM events according to the SHMEM specification. These relations arise from the added SHMEM operations and their observable accesses, similar to how the `so` and `hb` relations arise from added synchronization semantics in an SC for DRF model. Each relation formalizes a constraint described in Section II.C.3), and Figure 5 illustrates how these

```
Implicit local visibility (ilv)
ilv = (A, B): ∃op, B ∈ op.accesses
      ∧ (A, op) ∈ hb ∧ A.pe = B.pe = A.home
Local completion order (lco)
lco = (A, B): ∃op1∃op2, A ∈ op1.accesses
      ∧ op1.is_blocking ∧ A.pe = A.home
      ∧ ((op1, B) ∈ hb
      ∨ (∃op2, B ∈ op2.accesses
      ∧ (op1, op2) ∈ hb)
Remote delivery order (rdo)
rdo = (A, B) : ∃op1∃op2∃op_f,
     op_f.type == fence
     ∧ B ∈ op2.accesses ∧ (op_f, op2) ∈ hb
     ∧ ((A ∈ op1.accesses ∧ (op1, op_f) ∈ hb
     ∧ op1.ctx==op2.ctx==op_f.ctx ∧ A.pe == B.pe)
     ∨ (A, op_f) ∈ hb)
Remote completion order (rco)
rco_pre = (A, op1) :
      op1.type ∈ {quiet, barrier}
      ∧ ((A, op1) ∈ hb
      ∨ (∃op2, A ∈ op2.accesses
      ∧ (op2, op1) ∈ hb ∧ op1.ctx == op2.ctx))
rco_post = (op1, A) :
      op1.type ∈  {quiet, barrier}
      ∧ ((op1, A) ∈ hb
      ∨ (∃op2, A ∈ op2.accesses
      ∧ (op1, op2) ∈ hb ∧ op1.ctx == op2.ctx))
rco = rco_pre ∪ rco_post
API synchronization order (aso)
aso = (A, B) : ∃op1∃op2,
      A ∈ op1.accesses ∧ B ∈ op2.accesses
      ∧ (A, B) ∈ rf ∧ op1.ctx == op2.ctx
      ∧ op1.is_sync ∧ op2.is_sync
```

**Figure 4: API ordering relation formal definitions.**

relations are applied to an example SHMEM communication pattern. In this example execution, PE0 and PE1 communicate the data object `A` with an asynchronous `put_nbi` operation, using `fence`, `signal`, and `wait_until` operations to enforce ordering in the presence of asynchrony. Both PEs then use collective `barrier` operations to synchronize again. SHMEM operations (denoted by rounded boxes in Figure 5) are ordered within the calling thread by `po`, but their observable accesses are constrained *only by the added API relations* (in addition to `po` orderings internal to the operation). In addition, the paired barrier calls from each PE are fused into a single operation event, and the memory accesses of the bulk put transfer operation are fused into a single bulk read access and a single bulk write access. We next describe the relations depicted in Figure 5 and formalized in Figure 4.

**Implicit local visibility (`ilv`)**: We define the `ilv` relation to order the observable local accesses from a SHMEM operation with all prior accesses that are ordered before that operation by `hb` (`hb` is previously defined for DRF in Figure 2). This is formalized in Figure 4, where event `A` and `B` represent memory accesses, `op` represents a SHMEM operation, and `op.accesses` represents the set of observable accesses defined for `op`., and `home` and `pe` properties represent the issuing PE and target PE of each access. In Figure 5, `ilv` orders the store of `A` in PE0 (and all prior `hb`-ordered accesses) with the load of `A` from the put operation.

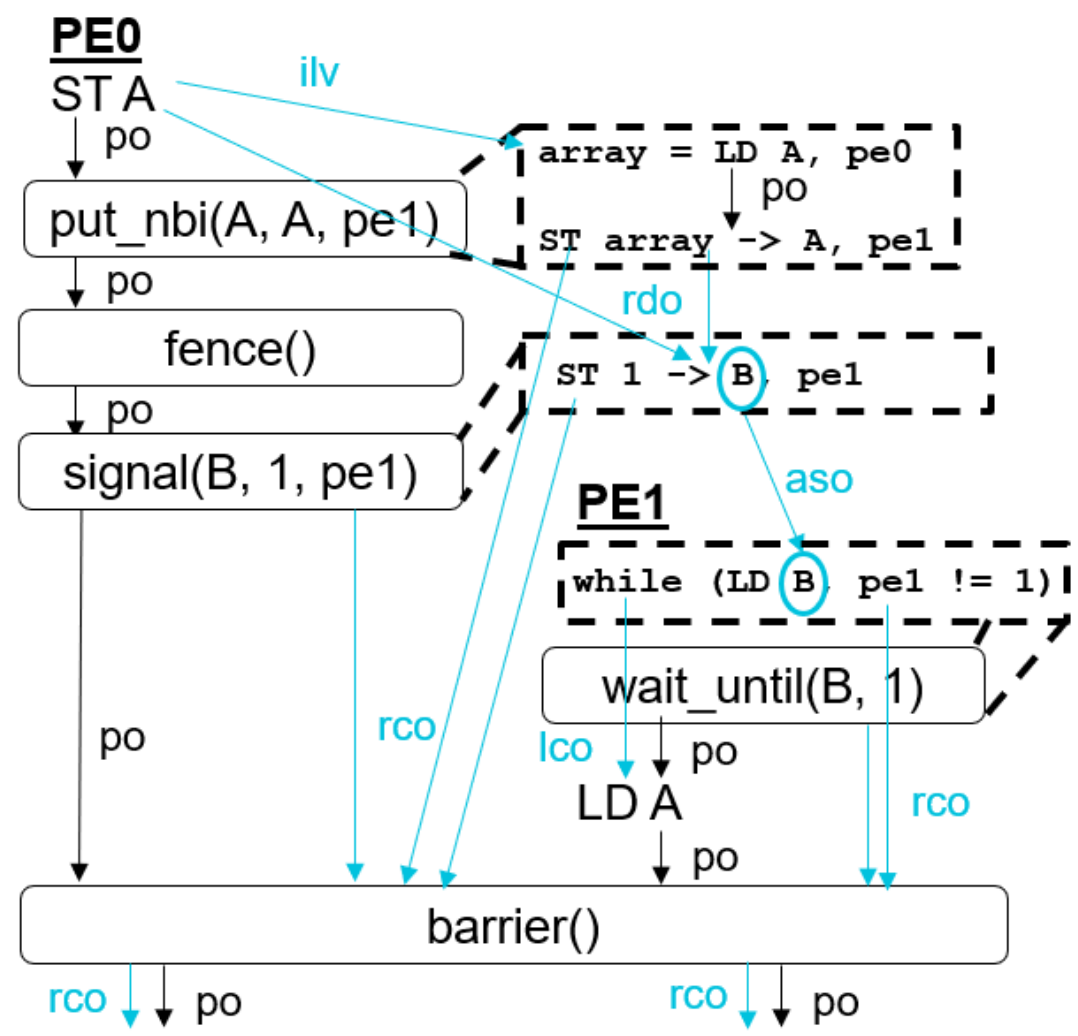

**Figure 5: Example SHMEM code implementing communication between a producer (PE0) and a consumer (PE1). Dotted boxes show observable accesses for each operation. Blue arrows represent added relations (some relations omitted for clarity).**

**Local completion order (`lco`)**: We define `lco` to order observable local accesses from a blocking SHMEM operation (and the operation itself) before all accesses that are ordered after the operation by `hb`. We formalize this relation in Figure 4, where `A` and `B` represent memory accesses, `op1` represents a SHMEM operation that issues `A`, and `B` is either `hb` ordered with `A`, or is issued from an operation (`op2`) that is `hb` ordered with `op1`. In Figure 5, `lco` orders the load of `a` in PE1 (and all subsequent `hb`-ordered accesses) with the load of `B` from the (blocking) `wait_until` operation.

**Remote delivery order (`rdo`)**: We define `rdo` to order accesses (`A`, `B`) if either 1) they are both issued by operations that share a context, target the same PE, and are separated by a fence (`op_f`) in `hb` order, or 2) `A` is a normal memory access that is `hb` ordered before an operation that issues `B`, and A is separated from the operation by a fence in `hb` order. We formalize this relation in Figure 4, where `op.ctx` represents the context associated with `op`. In Figure 5, `rdo` orders both the normal store of `A` from PE0 and the store of `A` from the `put` operation in PE0 with the store of `B` from the fence-separated `signal` operation.

**Remote completion order (`rco`)**: We define `rco` in Figure 4 to order `barrier` and `quiet` operations with all normal accesses and operation-issued accesses ordered before (`rco_pre`) or after (`rco_post`) the operation by `hb` (in the case of operation-issued accesses, the operation contexts must also match). In Figure 5, `rco` orders all remote accesses from `hb`-ordered SHMEM operations in PE0 and PE1 with the `barrier` operation spanning both PEs.

**API synchronization order (`aso`)**: We define `aso` in Figure 4 to order observable accesses that are related by `rf` in the candidate execution if they are both issued by synchronizing operations (op.is_sync returns true if op is a synchronizing operation) and share a context. In Figure 5, aso orders the store of B from the signal operation in PE0 with the consuming (rf-related) load of B from the wait_until operation in PE1.

*4) Updated Axiomatic Memory Model*

Next we leverage the new observable accesses and ordering relations to define the API-level DRF memory model, formalized in Figure 6. We first extend the hb relation of our baseline model to include the API ordering relations. The definition of a data race is expanded to ensure that conflicting accesses must also target the same PE, to classify a data race if a SHMEM observable access races with a normal access or with an access from a non-synchronizing operation, and to use the updated hb_api relation. Finally, we redefine sequential consistency for the API-level model (sc_api) as an acyclic constraint over po/rf/co/fr, but also over the added API orderings described in Section IV.B.3) (note that aso is implicitly included by rf). If every sc_api execution of a SHMEM program has no data_race_api, then that program is guaranteed to exhibit sc_api behavior.[1]

### C. Enforcing the API Memory Model

Adhering to the above SHMEM API-level model can be achieved by implementing SHMEM such that PL-level hb ordering is established for each added hb_api relation. This prevents reordering for relations that may participate in an hb_api chain, which includes ilv, lco, rdo, rco, and aso. Here we describe what is necessary for a generic HSA PL-level memory model where the SHMEM implementation may or may not offload SHMEM operations to a separate thread or hardware unit (referred to as the operation executor).

ilv and lco can be preserved by using acquire/release semantics to synchronize between the calling thread and the operation executor (required only if they are not the same thread). In addition, for blocking synchronization operations that read local memory, lco requires acquire semantics to be associated with the read since this can be used (along with aso) to establish an hb_api chain between synchronizing PEs. Preserving rdo and rco requires tracking the status of pending remote operations via a shared memory structure and blocking subsequent operations and/or accesses until prior operations reach a defined status (e.g., ordering on the network, or remote completion). They additionally require release semantics, since they may be used to establish hb_api ordering between a prior local access and a subsequent remote operation that targets the same variable. Preserving aso ordering requires that any observable accesses that may participate as a reader or writer in an aso relation use atomic semantics in the SHMEM implementation to avoid legally synchronizing API operations from introducing a PL-level data race.

```
API-level Happens-before
hb_api = (po ∪ so ∪ ilv ∪ lco ∪ rdo
    ∪ rco ∪ aso)+
API-level Data race
data_race_api = (A, B): A.addr == B.addr
  ∧ A.pe == B.pe
  ∧ ¬((A.atomic ∧ B.atomic)
  ∨ (A ∈ op_a ∧ B ∈ op_b
  ∧ op_a.is_sync ∧ op_b.is_sync))
  ∧ (A.type == ST ∨ B.type == ST)
  ∧ (A, B) ∉ hb_api ∧ (B, A) ∉ hb_api
API-level Sequential Consistency
sc_api = acyclic(po ∪ rf ∪ fr ∪ co ∪ ilv
    ∪ lco ∪ rdo ∪ rco)
```

**Figure 6: Updated API-level SC for DRF memory model formalism.** sync_types **refers to synchronization operation types defined in Table 1**

## V. Validating Model Formalism

We validate the API-level model described above by implementing it in Alloy [33] and verifying expected behavior for a range of litmus tests. This implementation is applied on top of a formalization of the LLVM compiler infrastructure's memory model[2] [34][37].

The litmus tests are based on a canonical set of memory ordering litmus tests summarized in Table 3. They have been modified to use non-synchronizing SHMEM calls (put, get) for non-synchronizing data accesses and synchronizing SHMEM calls (signal_set, atomic_rmw, atomic_fetch, wait_until) for synchronizing accesses. The minimal ordering is applied to ensure the unsatisfiability of the non-SC behavior. In addition, we generate variants of these tests in two ways: 1) We revert some of the API calls to normal memory accesses to test interaction of SHMEM semantics with PL-level ordering constraints. 2) We relax the ordering constraints in the tests in order to induce non-SC behavior and/or data races. This is accomplished by removing fences or downgrading API calls (synchronizing->non-synchronizing calls, quiet->fence, or lco-carrying local access -> non-lco remote access). The message passing litmus test and its variants are provided as examples in Figure 8.

Across all litmus tests, we successfully used Alloy to validate that the expected SC, non-SC, and racy behavior is observed under the formalized SHMEM memory model.

## VI. Optimizing the API-Level Memory Model: Targeted Ordering

In this section we describe an important optimization to device-side communication frameworks: targeted ordering for network data. We show that, with some minor changes to our formally defined API memory model, we can achieve the performance benefits of highly optimized code while retaining a mature, standardized, and precisely defined API. In Section VII, we evaluate the performance impact of targeted ordering

[1] Figure 5 illustrates how added SHMEM API relations order conflicting accesses to *a*, preventing a data race

[2] This real-world memory model differs slightly from the generic SC for DRF model presented above, but the extension to LLVM extends the hb relation in the exact same way.

```
(a) Original
PE0:ST 1 -> x
    ST_rel 1-> flag
PE1:r0 = LD_acq flag
    if r0==1:
      r1 = LD x

unsatisfiable: r0=1, r1=0
data race: no
```

```
(b) API variant
PE0:shmem_put(x, xsrc, pe1)
    shmem_fence()
    shmem_atomic_set(flag, 1, pe1)
PE1:wait_until(flag==1)
    get(y, x, pe1)

unsatisfiable: y!=xsrc
data race: no
```

```
(c) Mixed variant
PE0:ST 1 -> x;
    shmem_fence();
    shmem_signal_set(flag, 1, pe1);
PE1:wait_until(flag==1)
    r0 = LD x

unsatisfiable: r0=0
data race: no
```

```
(d) Non-SC variant
PE0:shmem_signal_set(x, 1, pe1);
    shmem_signal_set(flag, 1, pe1);
PE1:wait_until(flag==1)
    r0 = shmem_signal_fetch(x)

satisfiable: r0=0; data race: no
```

```
(e) Racy variant
PE0:shmem_put(x, xsrc PE1);
    shmem_signal_set(flag, 1, pe1);
PE1:wait_until(flag==1)
    get(y, x, pe1)

data race: yes
```

**Figure 8: Original message passing litmus test (a), the API version (b), and additional test variants using a mix of normal memory access and API calls (c), inducing non-SC behavior (d), and inducing a data race (e).**

$$\mathtt{ilv}_{\mathtt{com_id1}} = (\mathtt{A},\ \mathtt{B}) \cup (\mathtt{A},\ \mathtt{op}) \mid \exists \mathtt{op},\ \mathtt{B} \in \mathtt{op.accesses} \wedge (\mathtt{A},\ \mathtt{op}) \in \mathtt{hb}_{\mathtt{com_id1}} \wedge \mathtt{A.com_id} \le \mathtt{com_id1} \wedge \mathtt{com_id1} \le \mathtt{op.com_id}$$

$$\mathtt{hb_api} = \bigcup_{\mathtt{com_id} \in \mathtt{ids}} ((\mathtt{ppo} \cup \mathtt{so}_{\mathtt{com_id}} \cup \mathtt{ilv}_{\mathtt{com_id}} \cup \mathtt{lco}_{\mathtt{com_id}} \cup \mathtt{rdo}_{\mathtt{com_id}} \cup \mathtt{rco}_{\mathtt{com_id}} \cup \mathtt{aso}_{\mathtt{com_id}})^{+})$$

**Figure 7: Required updates to the `ilv` and `hb_api` relations for cache bypassing optimization. An update similar to the `ilv` addition is added to `lco`, `rdo`, and `rco`.**

across multiple workloads. While this section focuses on the SHMEM API, similar redefinition of other GPU communication libraries should obtain comparable improvements.

As discussed in Section IV.C, a GPU-based SHMEM implementation requires acquire and release semantics to preserve ordering for SHMEM operations and data accesses, which trigger cache flush and invalidate actions in GPUs. These actions add latency and unnecessarily degrade cache reuse for unrelated accesses. We propose a more targeted ordering semantic that avoids the need for flush and invalidate actions by only establishing ordering communicated data, and using cache bypassing when accessing this data.

This relaxation can be exposed at the API-level memory model through a new property we refer to as a communication ID, or com_id. A com_id is associated with each SHMEM operation[3] and with each memory access. Users may dynamically associate com_id with individual accesses and operations, or they may statically assign a com_id to a memory allocation such that all accesses to that allocation use the associated com_id. If not explicitly specified, a default global com_id is used.

Each of the API ordering relations described previously is defined per com_id. Each com_id-specific relation is relaxed to only order non-observable accesses if their com_id is <= the com_id of the inducing operation, and the com_id of the inducing operation is <= that of the relation (the global com_id is > all other com_ids). This is formalized in the updated `ilv` definition shown in **Figure 7** (changes highlighted in red, similarly applied to `lco`, `rdo`, and `rco`). The added requirements ensure that memory accesses prior to an SHMEM operation are only ordered if the access and operation use suitable com_ids. Operations with the default global com_id still order all prior operations. We then redefine hb_api to be a union over the transitive closure of each com_id-specific set of API relations. We also replace `po` with `ppo`, or preserved program order, which is defined to restrict po to orderings that are explicitly enforced by the architecture (i.e., accesses to the same address, or separated by acquire/release/fence semantics). These changes prevent transitivity from adding back excluded orderings.

**Table 3: Canonical litmus tests and the properties tested by each.**

| Litmus Test | Tested Property |
|---|---|
| Message Passing (MP) | Visibility through synchronization |
| Store Buffering (SB) | Store -> Load reordering |
| Load Buffering (LB) | Load -> Store reordering |
| Independent Reads Independent Writes (IRIW) | Multi-copy write atomicity |
| Write-to-Read Causality (WRC) | Does causality propagate through a read and dependent write? |
| Write-Write Coherence (CoWW) | Per-location SC |
| Transitive Synchronization (ISA2) | Transitivity of synchronization |

When applying acquire/release semantics to preserve these com_id-specific relations, only data with the target com_id requires ordering. If all data with the associated com_id is defined to be accessed via cache bypassing loads and stores, then these acquire and release operations do not need to flush or invalidate the cache at all.[4]

## VII. Applying the Targeted Ordering Optimization

We demonstrate the value of targeted ordering by applying it within two different SHMEM backend implementations in a production level SHMEM implementation (rocSHMEM) and by comparing performance with and without targeted ordering for three SHMEM workloads.

[3] This can be added as a field in the communication context, which is an existing object associated with each SHMEM operation.

[4] Unlike synchronization domains (described in Section III), this benefit can be leveraged by a single kernel.

### A. SHMEM Backend Modification: IPC and GDA

We implement the the cache bypassing optimization in two rocSHMEM backends: inter-process communication (IPC) backend for intra-node communication over xGMI, and the GPUDirect Async (GDA) backend for inter-node communication over InfiniBand. IPC is used for all workloads; GDA is only evaluated with DeepEP.

```
shmem_put_GDA(dst, src, size, dst_pe):
    ... //allocate and populate xfer packet
    //ring doorbell (store-release)
    cache_wb_l2 //sys-scope cache flush
    waitcnt //block until prior reqs complete
    ST_sys 1-> doorbell //store bypasses L2

shmem_signal_set_GDA(flag, val):
    //same as shmem_put, but different xfer type

shmem_put_IPC(dst, src, size, dst_pe):
    for idx in 0:size:
        r0 = LD src[idx]
        ST r0 -> dst[idx]

shmem_signal_set_IPC(flag, val):
    ST_sys val -> flag //store bypasses L2

shmem_fence():
    cache_wb_l2 //sys-scope cache flush
    waitcnt //block until prior reqs complete

shmem_wait_until(tgt, val):
    while(LD_sys tgt != val) ;//load bypasses L2
    waitcnt
    cache_inv_l2 //sys-scope cache inv
```

**Figure 9: Assembly-level pseudocode required for IPC and GDA implementations of SHMEM calls. Cache flush and invalidate actions highlighted in red. fence() and wait_until() are the same for both IPC and GDA.**

In the baseline IPC implementation, a `put` operation copies data between PEs via normal load and store operations since PEs share a global address space. GDA implements `put` and `signal` by enqueueing a request packet on a send queue and updating a doorbell variable to signal to the NIC that the request packet and input buffer are ready for transfer, using release semantics on the doorbell update to establish `ilv` ordering. Both IPC and GDA use system scope release semantics in `fence` operations to establish `rdo` ordering, and acquire semantics in synchronizing read operations like `wait_until` to establish `lco` ordering. These acquire and release semantics are lowered to L2 cache flush and invalidate operations, as well as `waitcnt` instructions (these block further access until prior memory operations have completed), as illustrated in Figure 9.

When we apply targeted ordering, the acquire and release semantics of the `put`, `fence`, and `wait_until` operations only need to order data with the specified com_id. We define the target com_id to apply specifically to communicated network data buffers, and we only use cache bypassing accesses for this data. When targeted ordering is enabled, the flush and invalidate associated with acquire and release semantics are removed; only the `waitcnt` instruction remains to block until any pending thread accesses have completed. This provides both a latency and a cache efficiency benefit.

### B. Targeted Ordering Workloads

We evaluate three workloads using the SHMEM backends described above with and without targeted ordering. xGMI connects GPUs within a node, and InfiniBand connects GPUs across different nodes. The workloads are reduce-scatter (RS), fused GEMM+reduce-scatter (GEMMRS), and DeepEP. All communication kernels are profiled running on a single compute node. DeepEP is additionally profiled running and across two compute nodes. Each compute node contains eight AMD Instinct™ MI300X GPUs. The IPC SHMEM backend is used for communication between GPUs on the same node, which are connected via xGMI. The GDA backend is used for communication between GPUs on different nodes (DeepEP only), which are connected via InfiniBand.

*a) Reduce Scatter (RS):* RS is single-kernel communication collective workload. It is commonly used in the ML and HPC domains, and exhibits high regularity, low cache reuse, and network-limited performance. Each GPU launches multiple workgroups, and in each iteration a workgroup transfers a vector of data to all other PEs, waits to receive a vector from each PE, and reduces the received vectors. We evaluate RS for a range of vector sizes with the IPC backend only, with each GPU transferring between 128KB and 256MB of data to remote PEs using a constant number of workgroups (128 workgroups, each composed of 512 threads).

*b) Fused GEMM+Reduce-Scatter (GEMMRS):* GEMMRS represents an additional level of workload complexity, fusing a GEMM compute kernel with a dependent reduce-scatter to achieve compute-communication overlap. GEMMRS exhibits high regularity, while cache reuse and the performance-limiting resource depend on the GEMM dimension. In GEMMRS, each GEMM output tile is transferred to a remote PE as soon as it is computed. As with RS, after transferring the chunks to the appropriate PEs, a workgroup waits until incoming transfers complete and reduces them with the local data for that PE's partition. Unlike RS, each workgroup operates on a fixed size output tile (128 elements), with the number of workgroups growing with the size of the output matrix. This avoids coarse-grain dependencies and allows network transfers from earlier workgroups to proceed concurrently with compute from later workgroups. The GEMM sizes are chosen to resemble common LLM decode workloads, with N and K dimensions (a proxy for LLM hidden dimensions) ranging from 4K-16K, and M dimension (a proxy for batch size) ranging from 1-256. While more complex than RS, GEMMRS is still a single kernel with a regular access and communication pattern.

*c) DeepEP:* DeepEP is a state-of-the-art communication library for expert parallelism, originally introduced in DeepSeek [1]. This workload exhibits irregularity (input-dependent token to expert routing), low cache reuse, and network-limited performance. It implements GPU kernels for the dispatch and combine stages surrounding the MoE feed-forward layer. DeepEP provides two kernel variants, optimized respectively for high throughput and low latency. Our study focuses on the low-latency variant, since the optimization considered here is more sensitive to communication latency. We use the embedding size associated with DeepSeek (7K) and we evaluate per-GPU token counts ranging from 64-256.

All of the evaluated workloads use `put`, `fence`, and `atomic_set` operations on the sender side and `wait_until` on the receiver side to implement inter-PE communication, and they can use targeted ordering with com_id's assigned to the data transferred by `put`, as described in Section VII.A.

## VIII. Targeted Ordering Performance Evaluation

Figure 10 plots the per-GPU network bandwidth achieved by RS with and without targeted ordering for the IPC backend. Targeted ordering improves performance by up to 2.48x for the IPC backend by avoiding the latency cost of cache flush operations. The greatest performance gains are at smaller transfer sizes where the relative impact of the flush latency is greater. At larger transfer sizes, performance is increasingly limited by network bandwidth rather than latency (both configurations saturate at around 220GB/s per GPU, which is roughly 50% of the theoretical peak), and the optimization benefit diminishes.

Figure 11 plots the relative speedup of targeted ordering for GEMMRS across a range of GEMM sizes with the IPC backend. At these sizes, GEMMRS performance is primarily limited by communication latency and/or memory bandwidth. The benefit of targeted ordering comes from avoiding flush latency AND avoiding the impact of flush and invalidate operations on cache reuse, since the SHMEM operations are executed concurrently with GEMM compute. As the M dimension grows, cache reuse potential for the GEMM operation increases (the KxN weight matrix is accessed M times), and more SHMEM transfers are necessary (the MxN output matrix is transferred in fixed tile sizes), so targeted ordering benefit increases. As N and K dimensions grow, the GEMM operation increasingly forms a performance bottleneck, so targeted ordering benefits are reduced. Overall, targeted ordering offers up to 1.77x speedup for GEMMRS on IPC.

Figure 12 plots the relative speedup of targeted ordering for DeepEP across a range of token counts with both IPC and GDA backends. Similar to RS, DeepEP is purely a communication workload, and the primary benefit of targeted ordering comes from avoiding flush latency rather than cache effects. Although the irregularity of DeepEP leads to some fluctuation in performance across token counts, targeted ordering consistently provides between 10-20% speedup relative an unoptimized baseline for most token counts. The largest gains (1.18x for both IPC and GDA) are seen at lower token counts; this is consistent with the bandwidth benefits seen for the simpler RS workload, and is explained by the diminishing benefits of latency saving as performance becomes more throughput limited. Note also that speedups drop off earlier for the GDA backend, which has a lower peak bandwidth that can be saturated earlier.

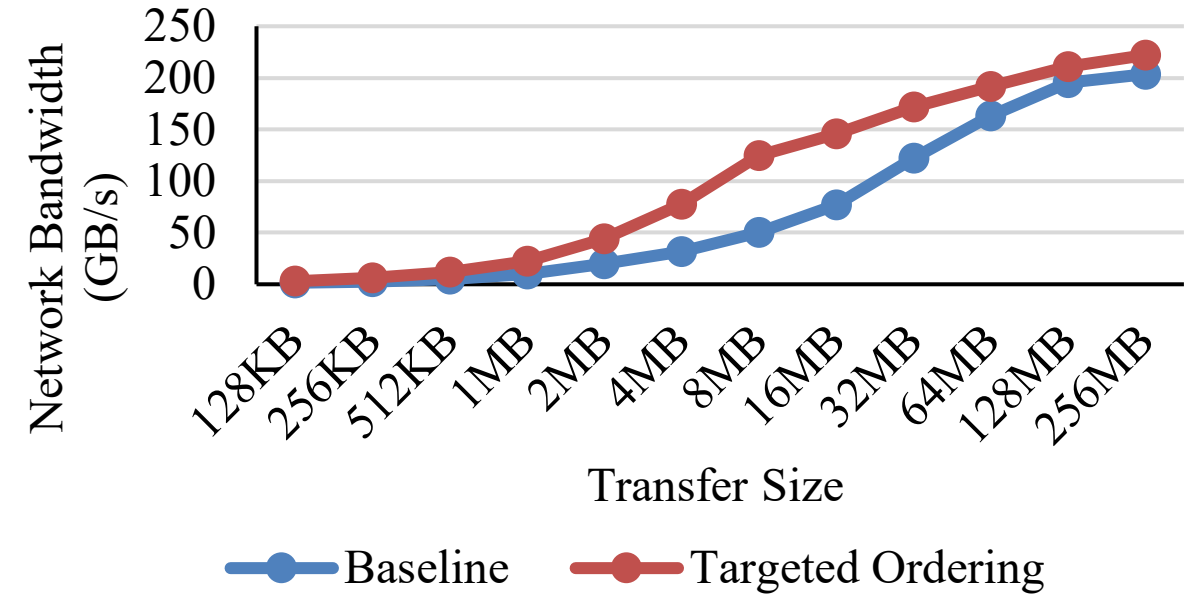

**Figure 10: Per-GPU xGMI bandwidth achieved with and without targeted ordering applied to reduce-scatter collective running 128 WGs per GPU, 512 threads per WG (IPC backend).**

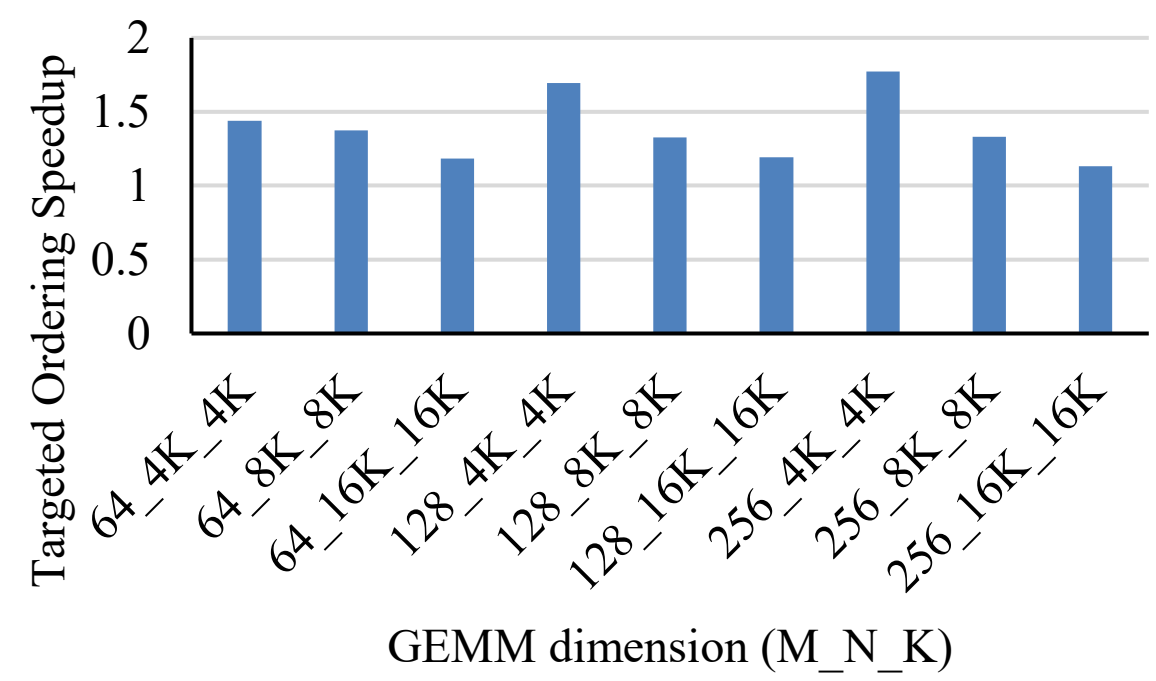

**Figure 11: Speedup achieved with targeted ordering applied to fused GEMMRS (IPC backend).**

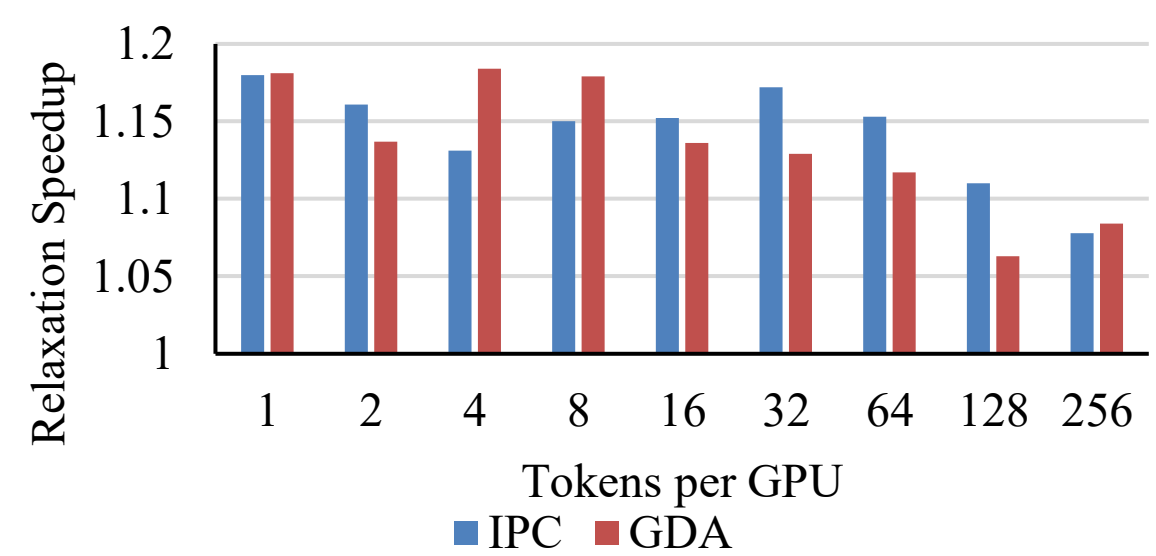

**Figure 12: Speedup achieved with targeted ordering applied to DeepEP (IPC and GDA backends).**

## IX. Conclusion

Communication APIs can simplify distributed programming for emerging ML workloads, but ambiguity in their specification leads to uncertainty for users and implementers. By formalizing an API-level SHMEM memory model, we provide a precise specification that enables safe optimization on either side of the API. We highlight this flexibility by formalizing the targeted ordering optimization for network data. Targeted ordering avoids costly flush and invalidate operations in SHMEM calls, leading to multiple sources of performance improvements (up to 2.48x) across multiple communication patterns and backends. By formalizing this optimization as an extension to the memory model, we make the optimization and associated ordering implications opt-in, and we ensure that users and implementers of the API agree on guaranteed behavior.

Looking beyond SHMEM, API-level memory models can be applied to other communication frameworks and interfaces,

many of which exhibit similar properties to SHMEM (i.e., coarse-grain, paired, and asynchronous operations). They can ease the introduction of new software optimizations as well as specialized hardware like data transfer engines and intelligent networking devices that may be used to accelerate communication in future systems. Distributed ML and HPC workloads increasingly rely on aggressive overlap of compute and communication tasks, and efficiency here requires that such interfaces are easy to understand, program, and optimize.

## X. ACKNOWLEDGEMENTS

Generative AI was used to help develop the alloy extensions for SHMEM formalism and associated litmus tests in this work.